\documentclass[
 reprint, superscriptaddress,
 amsmath,amssymb,
 aps,]{revtex4-2}

\usepackage{graphicx}
\usepackage{dcolumn}
\usepackage{bm}
\usepackage[usenames]{color}
\usepackage{colortbl}
\usepackage[dvipsnames]{xcolor}

\begin{document}

\preprint{APS/123-QED}

\title{Quantum Monte Carlo simulation of BEC-impurity tunneling}

\author{A.S. Popova}

 \email{chertkova.anastasia.phys@gmail.com}

 \affiliation{%
 	Russian Quantum Center, Bolshoy Bulvar 30, bld. 1,
 	Skolkovo, Moscow 	121205, Russia 
 }%

\author{V.V. Tiunova}
\affiliation{%
	Russian Quantum Center, Bolshoy Bulvar 30, bld. 1,
	Skolkovo, Moscow 	121205, Russia 
}%

\author{A.N. Rubtsov}%
\affiliation{%
	Russian Quantum Center, Bolshoy Bulvar 30, bld. 1,
	 Skolkovo, Moscow 121205, Russia 
}
\affiliation{%
 Department of Physics, Lomonosov Moscow State University,
  Leninskie gory 1, Moscow 119991, Russia
}

\date{\today}

\begin{abstract}
Polaron tunneling is a prominent example of a problem characterized by different energy scales, for which the standard quantum Monte Carlo methods face a slowdown problem. 
We propose a new quantum-tunneling Monte Carlo (QTMC) method which is free from this issue and can be used for a wide range of tunneling phenomena. We apply it to study an impurity interacting with a one-dimensional Bose-Einstein condensate and simultaneously trapped in an external double-well potential. Our scheme works for an arbitrary coupling between the particle and condensate and, at the same time, allows for an account of tunneling effects.  We discover two distinct quasi-particle peaks associated, respectively, with the phonon-assisted tunneling and the self-trapping of the impurity, which are in a crossover regime for the system modeled. 
We observe and analyze changes in the weights and spectral positions of the peaks (or, equally, effective masses of the quasi-particles) when the coupling strength is increased. Possible  experimental realizations using cold atoms are discussed.

\end{abstract}

\maketitle

\section{\label{sec:level1}Introduction}

The dynamics of a single mobile impurity interacting with a reservoir is one of the fundamental problems in condensed matter physics. 
The corresponding model, the  so-called polaron model,  was introduced to describe the coupling between electrons and lattice phonons in a dielectric crystal~\cite{landau1948effective}. Nowadays, the polaronic effects have been extensively  studied for the impurities in the Bose-Einstein condensates (BECs)~\cite{cucchietti2006strong, sacha2006self,   levinsen2015impurity,  grusdt2016tunable}, where tunable interaction between impurities and host atoms via the Feshbach resonance goes beyond the parameter range relevant for solids~\cite{drescher2019real}.  However, theoretical techniques based on perturbation  theory~\cite{rath2013field, christensen2015quasiparticle} and variational approaches~\cite{tempere2009feynman,casteels2012polaronic, grusdt2015renormalization, shchadilova2016quantum,volosniev2017analytical} result in different predictions at  the strong interacting even for the one-dimensional polarons~\cite{grusdt2017bose}. Nevertheless, this model became a sound benchmark for various many-body techniques~\cite{ drescher2019real, mistakidis2019effective}
with an unprecedented opportunity for their experimental testing~\cite{jorgensen2016,hu2016bose,ardila2016bose, mistakidis2020pump}. The transport of impurities  interacting with a many-body environment have been also investigated in optical lattices~\cite{bruderer2008transport,cai2010interaction, johnson2011impurity, theel2020manybodycollisional, palzer2009quantum}. Apart from that, progress in this area is also important for a deeper understanding of the  physics of neutral atoms in optical traps~\cite{frese2000single, bernien2017probing} and quantum dots~\cite{stauber2000electron, loss1998quantum} especially in the context of quantum information theory.

Generally, the incoherent tunneling effect~\cite{weiss1987incoherent,isakov2016understanding} with a nonlinear coupling is hard to study with analytical and numerical approaches. A tunneling particle interacting with a bath has at least two different energy scales -- a barrier height (related to the tunneling energy splitting) and the interaction strength. 
The path-integral quantum Monte Carlo (QMC) methods can be applied to this problem since they  are commonly  used to study quantum impurity models~\cite{lingua2018multiworm, ardila2015impurity}.  Moreover, the QMC studies of tunneling processes have gained an increased interest in the field of adiabatic quantum computing~\cite{inack2015simulated, stella2006monte}.  However, the straightforward application of the QMC algorithm for the simulation of tunneling is limited by their high computational complexity~\cite{ ardila2019analyzing, gull2011continuous, inack2018understanding, nemec2010diffusion}. This slowdown problem is related to a complex energy landscape for the Feynman path integral.

In this paper, we propose a special modification of the path integral QMC method for incoherent impurity tunneling and apply it to the polaron problem. We investigate tunneling of a single impurity immersed in a one-dimensional BEC and trapped in a double-well potential. 
Our method is based on splitting the path integral computation into two independent parts. The first one corresponds to the process of tunneling through a double-well barrier, which can be efficiently estimated,  and the second one  corresponds to a retarded interaction with the bath. 
 We consider a particle in a double-well potential in an numerically exact way, and BEC excitations are determined by integrating out the bosonic modes. 
The method relies on the assumption that the typical energy of the BEC-impurities interaction is comparable with the tunnel splitting, but much smaller than the barrier height. By performing an analytical continuation of the computed impurity's correlation functions, we calculate the density of states for different interaction strengths in the low-temperature limit. Moreover, we identify emergent peaks in the density of states as quasiparticle peaks and estimate their effective mass. Using the proposed QMC method, we  discover the crossover in the BEC-impurity system from phonon-assisted tunneling~\cite{oberli1990optical,vargas2020light} at a weak coupling to self-trapping in a strong interaction case~\cite{myasnikova1995band}.

The paper is organized as follows. In Sec. II we discuss the general formalism as well as the  model and the proposed method. Section III contains details of the proposed QMC scheme for the BEC-polaron tunneling in the case of the two-mode bath, comparison with the exact diagonalization method,  and results for a model with the continuous spectrum of the bath.  We provide conclusions in Sec. IV.
 
\section{\label{sec:level2}Model and method}

\subsection{\label{sec:level2a}Frohlich-Bogoliubov model}

 We use the Frohlich Hamiltonian, which describes the impurity in a Bose-Einstein condensate in the Bogoliubov approximation \cite{cucchietti2006strong, sacha2006self, tempere2009feynman}
\begin{align}\label{H0}
\hat{H}_{FB}=\frac{\hat{p}^2}{2m_{I}}+\sum_{ k\not =0}{\hbar\omega_{k}\hat{b}_{ k}^{\dagger}\hat{b}_{ k}}+
\sum_{k \not =0} \left(V_{ k}e^{-i  k \hat{ x}}\hat{b}_{-k}+h.c.\right)
\nonumber \\   V_{k}= \frac{a_{IB}\sqrt{n_{0}}}{\sqrt{2\pi} M}  \left[\frac{(\xi k)^2}{2+(\xi k)^2}\right]^{1/4} \,
 \omega_{k} = ck \left[1+\frac{(\xi k)^2}{2}\right]^{1/2}
\end{align}
where $\hat{p}$ and $\hat{x}$ are momentum and position operators of the impurity atom with mass $m_I$, 
$\hat{b}_{k}^{\dagger}$ is the creation operator of the Bogoliubov excitation with momentum 
$k$ and frequency $\omega_{k}$, $V_{k}$ -- interaction strength of phonon modes with the impurity atom, $c $ and $\xi$ are  the speed of sound in BEC and its healing length respectively,  $a_{IB}$ -- the boson-impurity scattering length,  $n_{0}$ is the BEC-density, and $M^{-1}={m_{B}}^{-1}+{m_{I}}^{-1}$ is the reduced mass, where $m_B$ is the mass of a host atom.
\par
 To study the equilibrium properties of the polaron tunneling  we consider the impurity in the double-well potential:
\begin{eqnarray}\label{H}
\hat{H} = \hat{H}_{FB} +  \kappa \left(-\frac{\hat{x}^2}{2} + \frac{\hat{x}^4}{4} \right)~
\end{eqnarray}
 where we set $\hbar$, $e$, $d$ to unity  ($e = c/\xi$ is the energy in the polaronic units \cite{grusdt2015renormalization}, $2d$ is the distance between wells of the double-well potential).

\begin{figure}[b]
	
	\begin{minipage}[h]{1\linewidth}
		\includegraphics[width=1\linewidth]{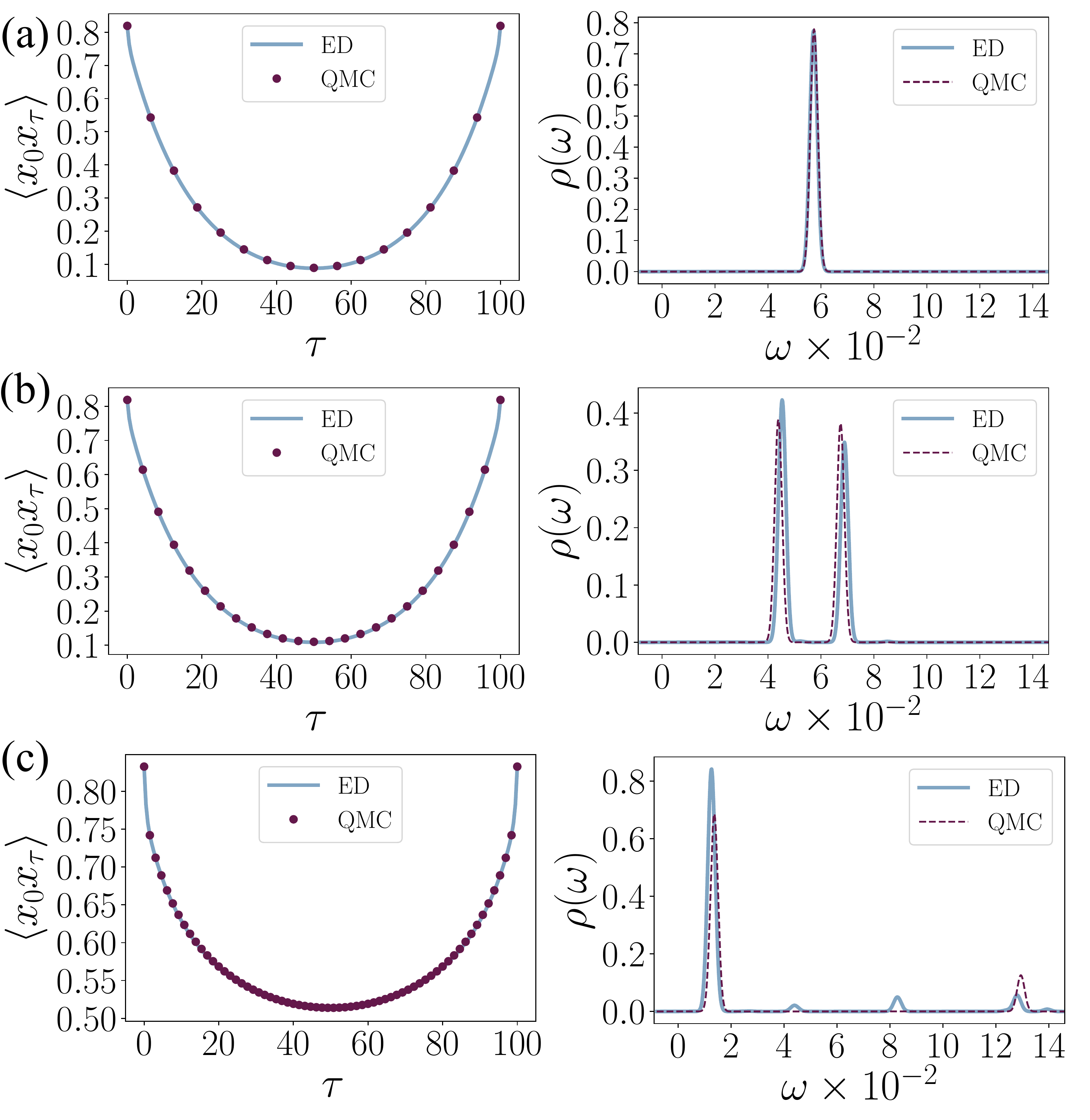}
	\end{minipage}

	\caption{Correlation functions~$ \langle x _{0} x_{\tau}\rangle $ and spectral densities of states ~$\rho(\omega)$ for $^{39}$K impurity, which interacts with the two resonance modes of Bogoliubov excitations~($k= \pm 0.805, \omega_k = \Delta E~\sim~0.057$) of $^{87}$Rb condensate (the healing  length~$\xi=2.5$) and trapped in a double-well potential with $\kappa = 10.24$ for different values of the scattering length:  (a)~$a_{IB} = 0.0$, (b)~$a_{IB} = 0.08$, (c)~$a_{IB} = 0.34$. Dotted and dashed lines correspond to the QTMC simulation with~$10^{10}$~steps, solid lines are the exact diagonalization results, $\beta~=~100$.
	}
	\label{fig:2}
\end{figure}

\subsection{\label{sec:level2b}Quantum-tunneling Monte Carlo method}

The Feynman path integral defines a transition amplitude as a sum of all possible paths between  given initial and final configurations of a quantum system.
The standard path integral quantum Monte Carlo method utilizes this idea in a sampling of a large number of discrete trajectories in imaginary time~\cite{westbroek2018user}. On every step of the algorithm, the current trajectory is changed according to the Metropolis condition~\cite{metropolis1949monte}. The more steps of the method are applied, the closer result becomes to the exact one. 
\par
 For tunneling problems  with multiple minima of the energy landscape the standard QMC sampling of valuable trajectories leads to the exponential growth  of computational time~\cite{gull2011continuous,parolini2019tunneling, inack2018understanding, nemec2010diffusion}. In other words, the standard path integral QMC cannot reach any stable result in a reasonable amount of time. The reason of the QMC scheme failure is that the tunneling time is much larger than the time scale of interaction. For BEC, this gives  $\hbar /\kappa\ll \hbar M/a_{IB}\sqrt{n_0}$, and as a consequence, the calculation requires a very fine grid for the trajectories. To overcome this problem, we proposed a new algorithm -- the quantum-tunneling Monte Carlo method (QTMC). Our approach separates the path-integral computation into two parts. The impurity tunneling in a double-well contribution is accounted through a numerically exact calculation of its propagator and the BEC-excitations are integrated out in the low-temperature limit, which results in the following retarded action

\begin{eqnarray}\label{Z}
Z = \int \mathcal{D}[x,b^{\dagger},b] e^{-S[x,b^{\dagger},b]} = \int \mathcal{D}[x] e^{-S_{I}[x]} e^{-S_R[x]}  \nonumber \\
e^{-S_R[x]} \equiv \int \mathcal{D}[b^{\dagger},b]  e^{-S_{B}[x, b^{\dagger},b]} ~~~~~~~~~~~~
\end{eqnarray}
 where $S_{I}[x]$ is the action of the impurity in a double-well potential, $S_{B}[x]$ is the  action related to the impurity in the BEC and $S_{R}[x]$ is the  retarded polaronic action:

\begin{align}\label{Sb}
S_{R} = - 2 \sum_{k\neq 0} V_{\bf k}^{2}  \sum_{\tau,\tau'}
e^{-ik(x(\tau)-x(\tau'))} \times ~~~~~~~~~~~~~ \nonumber\\ ~~~~~~~~~~~~~
 \times\frac {e^{-\omega_{k} ((\tau-\tau')+\beta\Theta(-(\tau-\tau'))} } {1 - e^{-\omega_{k} \beta}} \delta\tau \delta\tau'
\end{align}
where $\Theta(\tau)$ is the Heaviside step function, $\beta$ is an inverse temperature and $\delta \tau$ is a time-slice.
 \par 
 
The action of the impurity~$S_{I}[x]$ can be defined through the Feynman propagator -- the probability amplitude to find a particle at the position $x'$ from $x$ in the time interval~$\delta \tau$~\cite{feynman2010quantum}. We exactly diagonalize the Hamiltonian for the particle in the double-well  potential and obtain the eigenfunctions and eigenvalues $\phi_{i}, E_{i}$ to estimate propagator via
 \begin{eqnarray}\label{K}
 K(x,x',\delta \tau) = \sum_{i} \phi^{*}_{i}(x)\phi_{i}(x')e^{-E_{i}\delta \tau}
\end{eqnarray}
 
 We use the finite-difference method to compute the eigenfunctions and eigenvalues $\phi_{i}, E_{i}$.
 
Now we discuss the algorithm of the QTMC procedure in more details:

1) an explicit calculation of the propagator of the non-interacting impurity~$K(x',x,\tau)$ in a numerically exact way. 

 2) an numerical evaluation of the retarded action $S_{R}$ for different values of~$x(\tau)-x(\tau')$.
 
 3) a Monte Carlo sampling of the impurity's trajectories  using the calculated propagator~$K(x',x,\tau)$ and the retarded action~$S_{R}$.

 Thus, we reduce the initial many-body problem to a single particle problem with the effective retarded action, which includes correlations of all orders in a numerically exact way. Our QTMC scheme  samples the impurity trajectories in imaginary time with periodic boundary conditions on a coarse time grid with the step~$\delta\tau\sim\Delta E^{-1}$.

 We apply the QTMC algorithm to find the correlation functions $\langle x _{0} x_{\tau}\rangle $ of the impurity in the imaginary time. Using the QTMC data, we obtain a density of states~(DOS) on real axis through a Fourier transform~\cite{levy2017implementation,ghanem2016analytic}.
 \begin{equation} \label{ro2}
\langle x_0 x_{\tau}\rangle = \frac{1}{2\pi} \int_0^{\infty} \frac{e^{-\omega \tau} + e^{-\omega (\beta - \tau)}}{1-e^{-\omega \beta}}\rho(\omega) d\omega
\end{equation} 

The integral~(\ref{ro2}) is restricted to the positive frequencies~$\omega$, which is possible since~$\langle x_0 x_{\tau}\rangle=\langle x_0 x_{\beta-\tau}\rangle$~\cite{ghanem2016analytic}.

 We note that there are certain similarities between our approach and the strategy used in Refs.~\cite{pilati2006equation,ceperley1995path,krauth1996quantum} to simulate hard-core interactions in quantum bosonic fluids. In these works, the exact propagators for the single particle in a trap are also used for the effective sampling of configurations. However, we stress that Refs.~\cite{pilati2006equation,ceperley1995path,krauth1996quantum}~use the two-particle approximation for the density matrix, whereas our approach is formally numerically exact and preserves correlations at all orders.

\subsection{\label{sec:level2c} Maximum entropy method}

Generally, a finding the analytical continuation~$\rho(\omega)$  from the imaginary-time Green's function~$ \langle x _{0} x_{\tau}\rangle $ is  ill-conditioned problem \cite{yoon2018analytic}, i.e., the solution is highly sensitive to a noise of the input data. The maximum entropy method (MaxEnt) is a widely used approach to extract the analytical continuation~$\rho(\omega)$  from the correlation functions~$ \langle x _{0} x_{\tau}\rangle $.  The main idea of the MaxEnt method is to minimize the cost function $\frac{1}{2}\chi - \alpha S[\rho]$, where $\chi$ is the quadratic loss function of $\langle x_0 x_{\tau} \rangle$, $S[\rho]$ is  the Shannon entropy term, and $\alpha$ is a regularization
parameter \cite{ghanem2016analytic}. The distribution that maximizes the information entropy is the one that is statistically most favored \cite{mead1984maximum}. In other words, the less is known about a target spectral function~$\rho(\omega)$, the higher its Shannon entropy is.  The proper initial guess for the distribution -- the default model -- helps to reconstruct the unique analytical continuation.  Thus, solving the non-linear optimization problem, we find a finite, smooth and positive spectral function~$\rho(\omega)$ without overfitting the correlation function's noise \cite{levy2017implementation}.

\begin{figure}[b]
	
	\begin{minipage}[h]{1\linewidth}
		\includegraphics[width=1\linewidth]{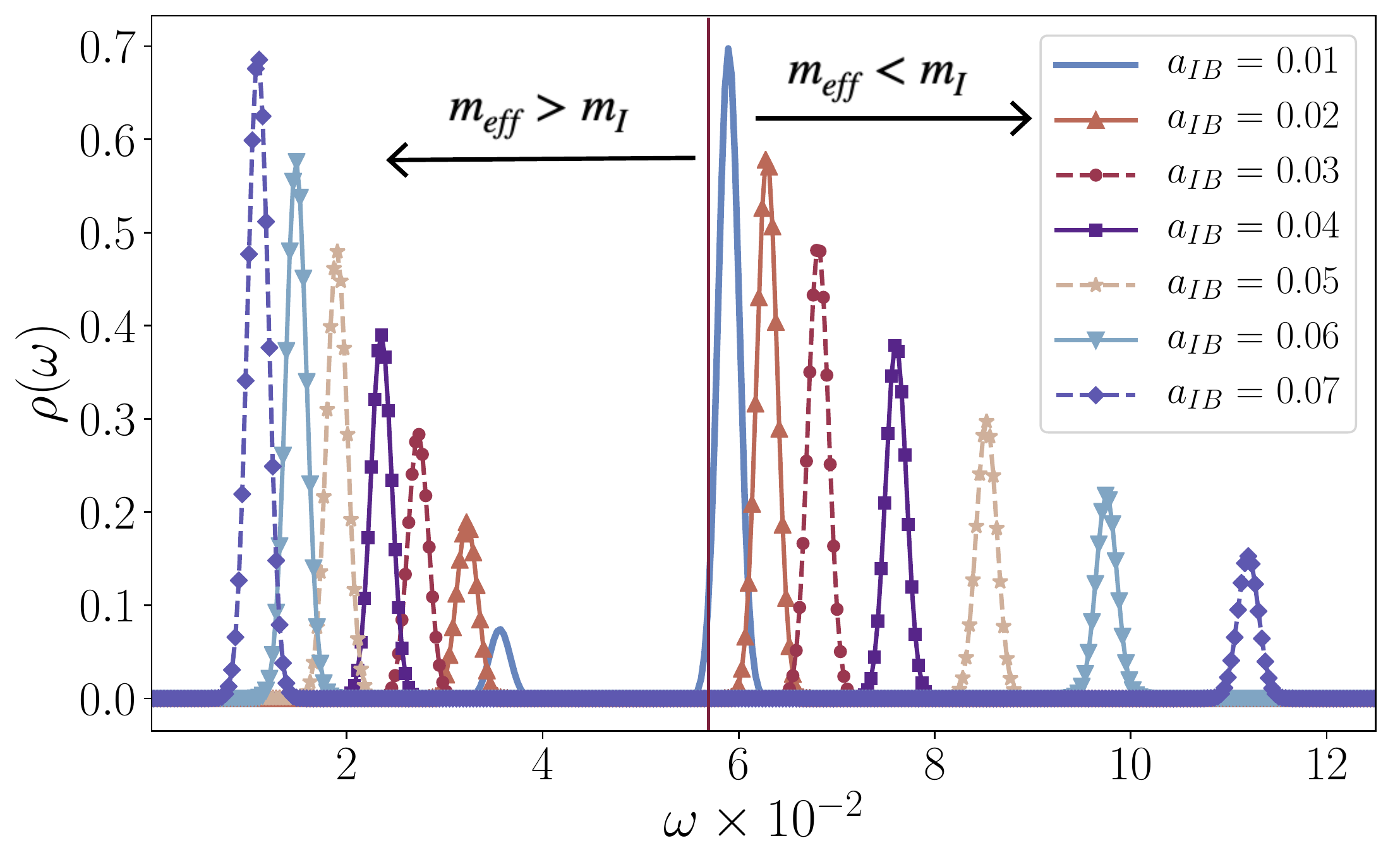}
	\end{minipage}

	\caption{Spectral densities of states~$\rho(\omega)$ for $^{39}$K tunneling impurity, which interacts with the continuous spectrum of Bogoliubov excitations in $^{87}$Rb condensate (the healing  length $\xi=2.5$),  $\kappa = 10.24$, $\beta=100$, $10^{10}$  QTMC steps. Vertical line corresponds to the tunneling energy splitting for zero interaction $a_{IB} = 0$.  }
	\label{fig:3}
\end{figure}

\section{\label{sec:level3}Results}

Here we present the numerical results for the system with the parameters corresponding to the $^{39}$K tunneling impurity in the $^{87}$Rb condensate. First, we benchmark the proposed quantum-tunneling Monte Carlo algorithm on the model~(\ref{H}) with the two resonance modes of Bogoliubov excitations. We solve this problem by the exact diagonalization (ED) method and the QTMC scheme (with 10 eigenfunctions $\phi_i$ held in Eq.~\ref{K})    for the different coupling strengths or equivalently the boson-impurity scattering lengths~$a_{IB}$~(see~Fig.~\ref{fig:2}).  The exact diagonalization algorithm provides the solution as a set of delta-function peaks for this problem.  In Fig.~\ref{fig:2} these peaks are slightly broaden for easier visualization.  The QTMC correlation functions~$ \langle x_0 x _ {\tau} \rangle $  were transformed  into  the density of states by the MaxEnt method.   This procedure approximates the DOS by  Gaussian peaks of a fixed width, defined by the accuracy  of QTMC calculations. Nevertheless, significant features of the density of states, namely a peak position and its amplitude, can be obtained without any restrictions on the boson-impurity scattering length $a_{IB}$. We note that the calculation time slightly grows with the coupling $a_{IB}$  since it is necessary  to use a more fine time grid in the strong coupling regime.

\par

For a free impurity in the double-well potential, we find a single peak in the DOS, corresponding to the tunneling splitting~$\Delta E$~(Fig.\ref{fig:2},~a). At the small coupling, this peak splits into two ones~(Fig.\ref{fig:2},~b). As interaction grows, the right peak shifts to the higher frequencies and its amplitude decreases, while the left DOS peak grows up and shifts to the lower frequencies. Finally, the lower-energy peak dominates for strong coupling, which might indicate a self-trapping of the impurity~(Fig.\ref{fig:2},~c).  There are slight differences in~$\rho(\omega)$ between the ED and QTMC peaks due to the finite accuracy of the MaxEnt procedure (see Section~\ref{sec:level2c} for details). Also, our scheme cannot resolve the smallest DOS peaks for the strong coupling~(Fig.\ref{fig:2},~c),  but we believe that these peaks do not significantly influence the impurity tunneling.

\begin{figure}[t]
	
	\begin{minipage}[h]{1\linewidth}
		\includegraphics[width=1\linewidth]{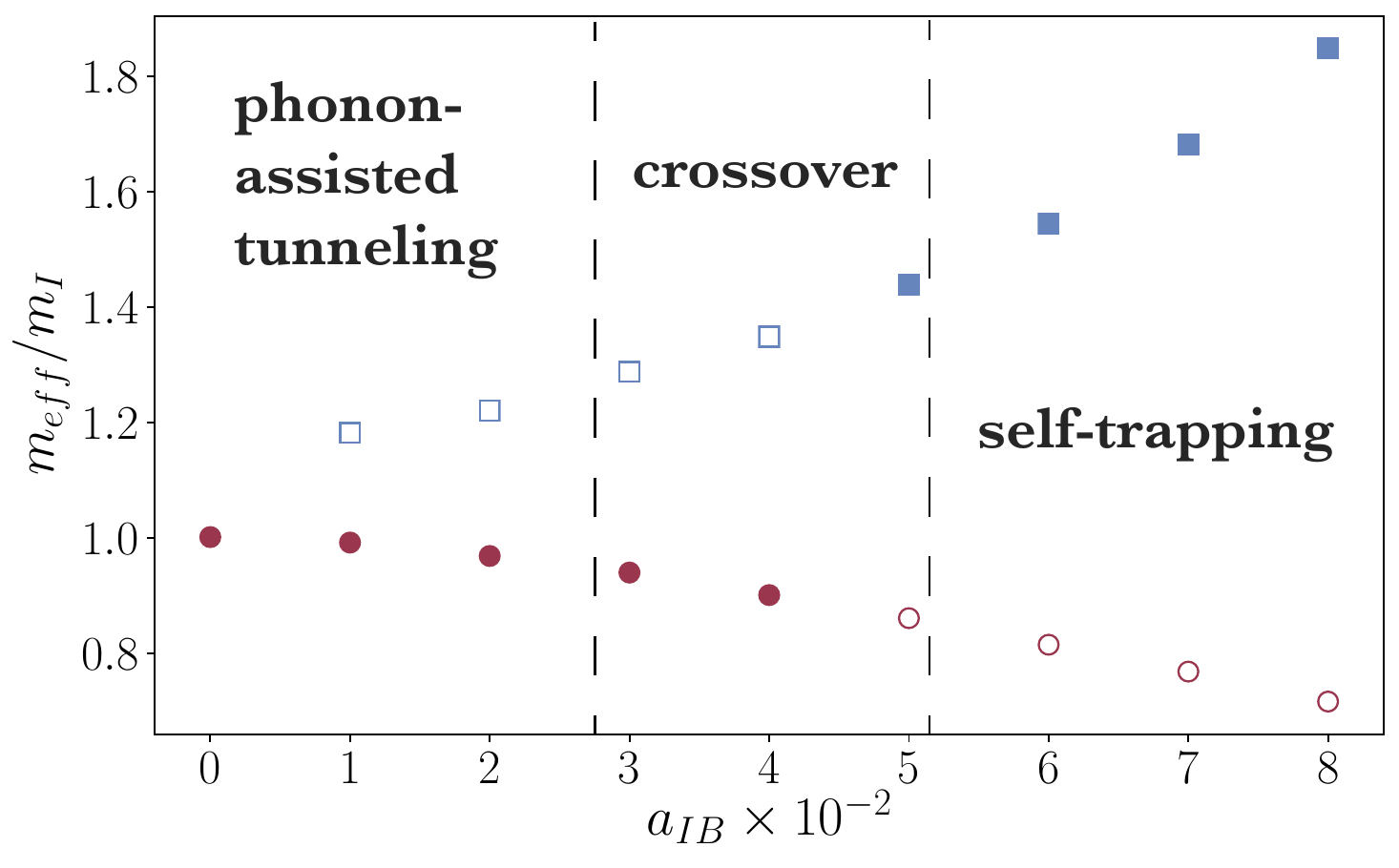}
	\end{minipage}

	\caption{Effective mass corresponding to the quasiparticle peaks of the density of states for~$^{39}$K tunneling impurity in~$^{87}$Rb condensate (the healing  length~$\xi=2.5$); circles represent the higher frequency peaks, squares -- the  lower frequency ones; solid (hollow) marks are dominant (lesser) peaks; $\kappa = 10.24$, $\beta=100$.}
	\label{fig:4}
\end{figure}

\begin{figure}[t]
	
	\begin{minipage}[h]{1\linewidth}
		\includegraphics[width=1\linewidth]{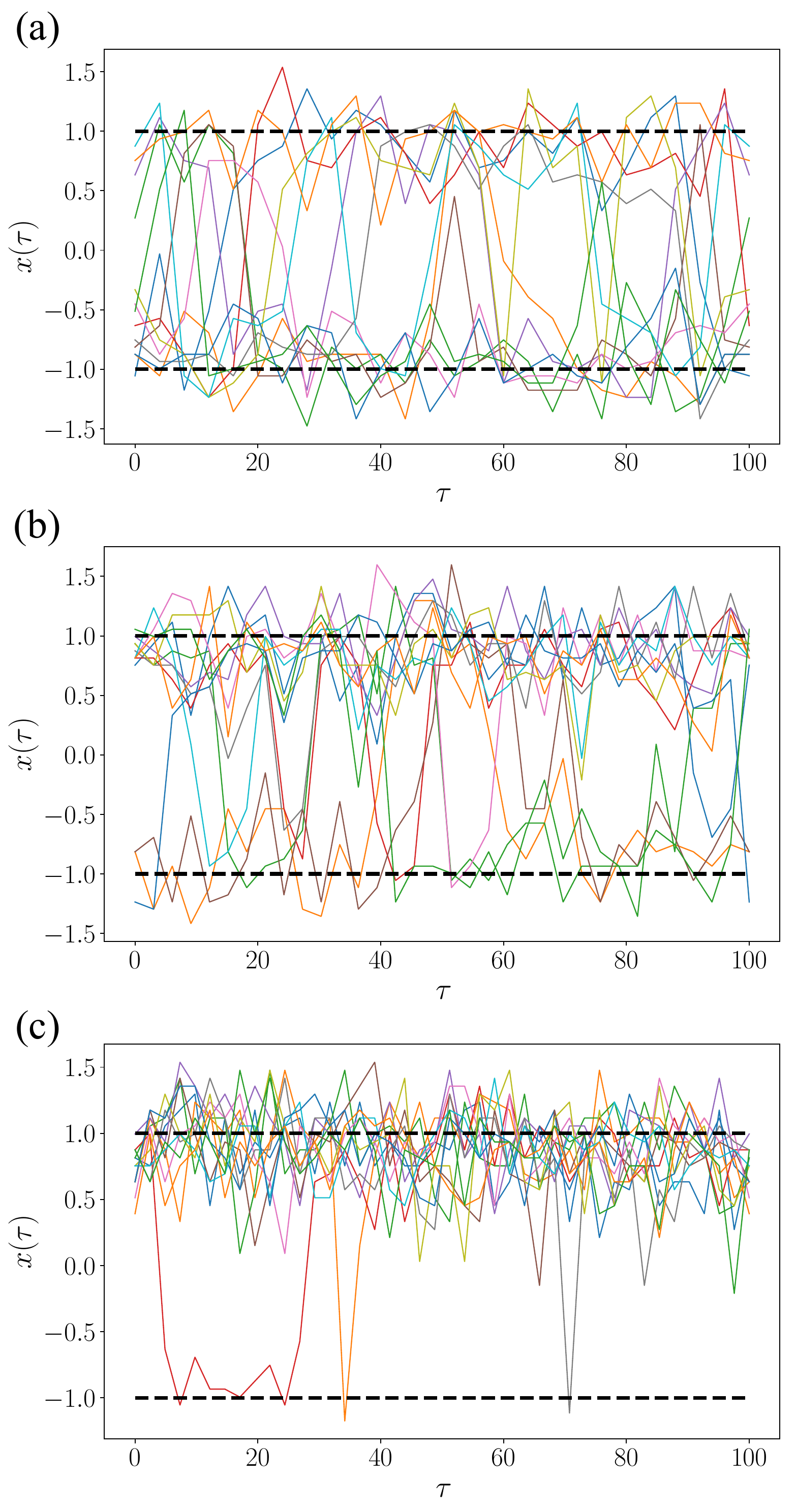}
	\end{minipage}

	\caption{Impurity trajectories~$x(\tau)$ in imaginary time for different values of interaction strength  with~$^{87}$Rb condensate (the healing  length~$\xi=2.5$): (a)~$a_{IB}~=~0.01$, (b)~$a_{IB}~=~0.05$, (c)~$a_{IB} = 0.09$; dashed lines correspond to the minima of the double-well potential, $\kappa~=~10.24$,  $\beta~=~100$. }
	\label{fig:5}
\end{figure}

Now let us discuss the QTMC results for the model~(\ref{H}) of the~$^{39}$K tunneling impurity with the continuous bosonic spectrum of Bogoliubov excitations (the healing  length~$\xi=2.5$,  the inverse temperature~$\beta = 100$ corresponding to~$T=160$~nK), which does not allow the ED treatment. Fig.~\ref{fig:3} shows the obtained spectral density of the states from the QTMC data. The spectrum of the bosonic modes was cut off at~$k_{max}=1.5$, which corresponds to the frequency cut~$\omega(k_{max})=3\Delta E$. We checked that obtained results do not depend on the specific choice of the cut.  The DOS behaviour is qualitatively similar to that of the two-mode problem. There is a single DOS peak for zero coupling, which is  related to the energy level splitting for the particle  in a double-well potential~$\omega_0  = \Delta E$ (vertical line in Fig.~\ref{fig:3}). In other words, a particle is tunneling from one well of the potential to another for tunneling time~$t= \pi/\Delta E$. For a small interaction strength (scattering length~$ a_ {IB} < 0.04 $), the tunneling DOS peak shifts to higher frequencies, and another small DOS-peak appears. The amplitude of the right DOS-peak decreases with coupling, but its position continues to move to higher frequencies, which means the decreasing impurity tunneling time in the presence of resonant phonons, i.e., phonon-assisted tunneling.  Simultaneously, the left DOS-peak grows and shifts to  lower frequencies.   For  the scattering length $a_{IB}>0.06$,   this peak becomes a dominating feature of the DOS, i.e., the tunneling time grows (Fig.~\ref{fig:3}).  It means that the impurity does not transfer into another well of potential or, in other words, it is self-trapped in the strong interaction regime. This process is interpreted as the formation of a heavy phonon cloud around the particle at which a bound state of the impurity emerges. 
\par
 For the obtained QTMC data, the  MaxEnt method does not signal any change in the widths of the DOS peaks. Since the DOS peaks are narrow, we can interpret them as the quasiparticle peaks and define their effective masses as a functions of scattering length (Fig.~\ref{fig:4}). For a given double-well potential (with $\kappa = 10.24$ in our case), we can employ the exact diagonalization to obtain a dependence of the tunneling splitting~$\Delta E$ on a particle mass. This dependence can be used to estimate an effective mass for each DOS peak.  We see that for a small interaction strength, which we refer to as a phonon-assisted tunneling region, the effective mass decreases (circles in Fig.~\ref{fig:4}).  For the large coupling, we observe an increase of the effective mass, which indicates that the impurity is localized in this regime (squares in Fig.~\ref{fig:4}). In the intermediate case, there is a crossover region, where two DOS peaks have nearly the same amplitude, and these two phenomena coexist. We emphasize that the defined effective masses  are not directly related to the impurity effective mass in the BEC in absence of the tunneling potential~\cite{ardila2015impurity} and further investigation is needed to elaborate on their connection.
 \par
In Fig.~\ref{fig:5} we also show the sampled  imaginary-time trajectories for the different scattering lengths $a_{IB}$. For a small interaction strength, the impurity tunnels freely from one well to another~(Fig.~\ref{fig:5},~a).
In the crossover region,  the impurity tunneling starts to lessen,  which results in a slightly asymmetric probability distribution~(Fig.~\ref{fig:5},~b). For a strong coupling, the tunneling process is almost suppressed, which  indicates  the self-trapping of impurity in one of the wells~(Fig.~\ref{fig:5},~c).

\section{\label{sec:level4}Conclusions}
In the present work we propose a special modification of the quantum Monte Carlo method for the tunneling problems (QTMC) in an external  environment, which can simulate equilibrium dynamics beyond perturbation theory. The QTMC scheme  enables us to investigate how the bosonic bath affects tunneling without an exponential slowdown usual for the standard QMC method. We apply this approach to study the tunneling of the BEC-impurity in one dimension in the Frohlich-Bogolubov approximation.  The impurity's correlation functions were calculated by the QTMC scheme and transformed into the density of states using the MaxEnt procedure. We verified our method  by employing exact diagonalization for the model bath with two resonance modes of the bosonic field. 
\par
For the continuous bosonic spectrum, we found that the BEC-impurity undergoes a crossover between the phonon-assisted tunneling at the weak coupling and self-trapping for the strong interaction. Also, we found the quasiparticle peaks in DOS and estimate their effective mass. These phenomena might be observed in the recent experiment realization \cite{jorgensen2016, hu2016bose} with an addition of two close harmonic optical dipole traps for the impurity \cite{spethmann2012dynamics, catani2012quantum}. 
Spectral response of the tunneling impurity in the BEC on radio-frequency pulses might be used for the observation of the crossover between the phonon-assisted tunneling and the self-trapping regimes. 
Moreover, it was shown that the inhomogeneous BEC could produce the effective double-well potential for the impurity during quench dynamics \cite{mistakidis2020many, mistakidis2019quench}. These works motivate a further study of the dynamical phenomena in the  explored BEC-impurity model and its extensions.

\subsection*{\label{sec:level}Acknowledgments}
The authors thank E.A. Polyakov for useful discussions. A.S.P. acknowledges the support by Theoretical Physics and Mathematics Advancement Foundation  ``BASIS'' through Grant No. 19-2-6-241-1.  V.V.T. acknowledges the support by Russian Science Foundation through Grant No. 19-71-10092.

\bibliography{polaron}

%apsrev4-2.bst 2019-01-14 (MD) hand-edited version of apsrev4-1.bst
%Control: key (0)
%Control: author (8) initials jnrlst
%Control: editor formatted (1) identically to author
%Control: production of article title (0) allowed
%Control: page (0) single
%Control: year (1) truncated
%Control: production of eprint (0) enabled
\providecommand{\noopsort}[1]{}\providecommand{\singleletter}[1]{#1}%
\begin{thebibliography}{56}%
\makeatletter
\providecommand \@ifxundefined [1]{%
 \@ifx{#1\undefined}
}%
\providecommand \@ifnum [1]{%
 \ifnum #1\expandafter \@firstoftwo
 \else \expandafter \@secondoftwo
 \fi
}%
\providecommand \@ifx [1]{%
 \ifx #1\expandafter \@firstoftwo
 \else \expandafter \@secondoftwo
 \fi
}%
\providecommand \natexlab [1]{#1}%
\providecommand \enquote  [1]{``#1''}%
\providecommand \bibnamefont  [1]{#1}%
\providecommand \bibfnamefont [1]{#1}%
\providecommand \citenamefont [1]{#1}%
\providecommand \href@noop [0]{\@secondoftwo}%
\providecommand \href [0]{\begingroup \@sanitize@url \@href}%
\providecommand \@href[1]{\@@startlink{#1}\@@href}%
\providecommand \@@href[1]{\endgroup#1\@@endlink}%
\providecommand \@sanitize@url [0]{\catcode `\\12\catcode `\$12\catcode
  `\&12\catcode `\#12\catcode `\^12\catcode `\_12\catcode `\%12\relax}%
\providecommand \@@startlink[1]{}%
\providecommand \@@endlink[0]{}%
\providecommand \url  [0]{\begingroup\@sanitize@url \@url }%
\providecommand \@url [1]{\endgroup\@href {#1}{\urlprefix }}%
\providecommand \urlprefix  [0]{URL }%
\providecommand \Eprint [0]{\href }%
\providecommand \doibase [0]{https://doi.org/}%
\providecommand \selectlanguage [0]{\@gobble}%
\providecommand \bibinfo  [0]{\@secondoftwo}%
\providecommand \bibfield  [0]{\@secondoftwo}%
\providecommand \translation [1]{[#1]}%
\providecommand \BibitemOpen [0]{}%
\providecommand \bibitemStop [0]{}%
\providecommand \bibitemNoStop [0]{.\EOS\space}%
\providecommand \EOS [0]{\spacefactor3000\relax}%
\providecommand \BibitemShut  [1]{\csname bibitem#1\endcsname}%
\let\auto@bib@innerbib\@empty
%</preamble>
\bibitem [{\citenamefont {Landau}\ and\ \citenamefont
  {Pekar}(1948)}]{landau1948effective}%
  \BibitemOpen
  \bibfield  {author} {\bibinfo {author} {\bibfnamefont {L.}~\bibnamefont
  {Landau}}\ and\ \bibinfo {author} {\bibfnamefont {S.}~\bibnamefont {Pekar}},\
  }\bibfield  {title} {\bibinfo {title} {Effective mass of a polaron},\
  }\href@noop {} {\bibfield  {journal} {\bibinfo  {journal} {Zh. Eksp. Teor.
  Fiz}\ }\textbf {\bibinfo {volume} {18}},\ \bibinfo {pages} {419} (\bibinfo
  {year} {1948})}\BibitemShut {NoStop}%
\bibitem [{\citenamefont {Cucchietti}\ and\ \citenamefont
  {Timmermans}(2006)}]{cucchietti2006strong}%
  \BibitemOpen
  \bibfield  {author} {\bibinfo {author} {\bibfnamefont {F.~M.}\ \bibnamefont
  {Cucchietti}}\ and\ \bibinfo {author} {\bibfnamefont {E.}~\bibnamefont
  {Timmermans}},\ }\bibfield  {title} {\bibinfo {title} {Strong-coupling
  polarons in dilute gas bose-einstein condensates},\ }\href@noop {} {\bibfield
   {journal} {\bibinfo  {journal} {Phys.\ Rev.\ Lett.}\ }\textbf {\bibinfo
  {volume} {96}},\ \bibinfo {pages} {210401} (\bibinfo {year}
  {2006})}\BibitemShut {NoStop}%
\bibitem [{\citenamefont {Sacha}\ and\ \citenamefont
  {Timmermans}(2006)}]{sacha2006self}%
  \BibitemOpen
  \bibfield  {author} {\bibinfo {author} {\bibfnamefont {K.}~\bibnamefont
  {Sacha}}\ and\ \bibinfo {author} {\bibfnamefont {E.}~\bibnamefont
  {Timmermans}},\ }\bibfield  {title} {\bibinfo {title} {Self-localized
  impurities embedded in a one-dimensional bose-einstein condensate and their
  quantum fluctuations},\ }\href@noop {} {\bibfield  {journal} {\bibinfo
  {journal} {Physical Review A}\ }\textbf {\bibinfo {volume} {73}},\ \bibinfo
  {pages} {063604} (\bibinfo {year} {2006})}\BibitemShut {NoStop}%
\bibitem [{\citenamefont {Levinsen}\ \emph {et~al.}(2015)\citenamefont
  {Levinsen}, \citenamefont {Parish},\ and\ \citenamefont
  {Bruun}}]{levinsen2015impurity}%
  \BibitemOpen
  \bibfield  {author} {\bibinfo {author} {\bibfnamefont {J.}~\bibnamefont
  {Levinsen}}, \bibinfo {author} {\bibfnamefont {M.~M.}\ \bibnamefont
  {Parish}},\ and\ \bibinfo {author} {\bibfnamefont {G.~M.}\ \bibnamefont
  {Bruun}},\ }\bibfield  {title} {\bibinfo {title} {Impurity in a bose-einstein
  condensate and the efimov effect},\ }\href@noop {} {\bibfield  {journal}
  {\bibinfo  {journal} {Physical Review Letters}\ }\textbf {\bibinfo {volume}
  {115}},\ \bibinfo {pages} {125302} (\bibinfo {year} {2015})}\BibitemShut
  {NoStop}%
\bibitem [{\citenamefont {Grusdt}\ and\ \citenamefont
  {Fleischhauer}(2016)}]{grusdt2016tunable}%
  \BibitemOpen
  \bibfield  {author} {\bibinfo {author} {\bibfnamefont {F.}~\bibnamefont
  {Grusdt}}\ and\ \bibinfo {author} {\bibfnamefont {M.}~\bibnamefont
  {Fleischhauer}},\ }\bibfield  {title} {\bibinfo {title} {Tunable polarons of
  slow-light polaritons in a two-dimensional bose-einstein condensate},\
  }\href@noop {} {\bibfield  {journal} {\bibinfo  {journal} {Physical review
  letters}\ }\textbf {\bibinfo {volume} {116}},\ \bibinfo {pages} {053602}
  (\bibinfo {year} {2016})}\BibitemShut {NoStop}%
\bibitem [{\citenamefont {Drescher}\ \emph {et~al.}(2019)\citenamefont
  {Drescher}, \citenamefont {Salmhofer},\ and\ \citenamefont
  {Enss}}]{drescher2019real}%
  \BibitemOpen
  \bibfield  {author} {\bibinfo {author} {\bibfnamefont {M.}~\bibnamefont
  {Drescher}}, \bibinfo {author} {\bibfnamefont {M.}~\bibnamefont
  {Salmhofer}},\ and\ \bibinfo {author} {\bibfnamefont {T.}~\bibnamefont
  {Enss}},\ }\bibfield  {title} {\bibinfo {title} {Real-space dynamics of
  attractive and repulsive polarons in bose-einstein condensates},\ }\href@noop
  {} {\bibfield  {journal} {\bibinfo  {journal} {Physical Review A}\ }\textbf
  {\bibinfo {volume} {99}},\ \bibinfo {pages} {023601} (\bibinfo {year}
  {2019})}\BibitemShut {NoStop}%
\bibitem [{\citenamefont {Rath}\ and\ \citenamefont
  {Schmidt}(2013)}]{rath2013field}%
  \BibitemOpen
  \bibfield  {author} {\bibinfo {author} {\bibfnamefont {S.~P.}\ \bibnamefont
  {Rath}}\ and\ \bibinfo {author} {\bibfnamefont {R.}~\bibnamefont {Schmidt}},\
  }\bibfield  {title} {\bibinfo {title} {Field-theoretical study of the bose
  polaron},\ }\href@noop {} {\bibfield  {journal} {\bibinfo  {journal}
  {Physical Review A}\ }\textbf {\bibinfo {volume} {88}},\ \bibinfo {pages}
  {053632} (\bibinfo {year} {2013})}\BibitemShut {NoStop}%
\bibitem [{\citenamefont {Christensen}\ \emph {et~al.}(2015)\citenamefont
  {Christensen}, \citenamefont {Levinsen},\ and\ \citenamefont
  {Bruun}}]{christensen2015quasiparticle}%
  \BibitemOpen
  \bibfield  {author} {\bibinfo {author} {\bibfnamefont {R.~S.}\ \bibnamefont
  {Christensen}}, \bibinfo {author} {\bibfnamefont {J.}~\bibnamefont
  {Levinsen}},\ and\ \bibinfo {author} {\bibfnamefont {G.~M.}\ \bibnamefont
  {Bruun}},\ }\bibfield  {title} {\bibinfo {title} {Quasiparticle properties of
  a mobile impurity in a bose-einstein condensate},\ }\href@noop {} {\bibfield
  {journal} {\bibinfo  {journal} {Physical review letters}\ }\textbf {\bibinfo
  {volume} {115}},\ \bibinfo {pages} {160401} (\bibinfo {year}
  {2015})}\BibitemShut {NoStop}%
\bibitem [{\citenamefont {Tempere}\ \emph {et~al.}(2009)\citenamefont
  {Tempere}, \citenamefont {Casteels}, \citenamefont {Oberthaler},
  \citenamefont {Knoop}, \citenamefont {Timmermans},\ and\ \citenamefont
  {Devreese}}]{tempere2009feynman}%
  \BibitemOpen
  \bibfield  {author} {\bibinfo {author} {\bibfnamefont {J.}~\bibnamefont
  {Tempere}}, \bibinfo {author} {\bibfnamefont {W.}~\bibnamefont {Casteels}},
  \bibinfo {author} {\bibfnamefont {M.~K.}\ \bibnamefont {Oberthaler}},
  \bibinfo {author} {\bibfnamefont {S.}~\bibnamefont {Knoop}}, \bibinfo
  {author} {\bibfnamefont {E.}~\bibnamefont {Timmermans}},\ and\ \bibinfo
  {author} {\bibfnamefont {J.~T.}\ \bibnamefont {Devreese}},\ }\bibfield
  {title} {\bibinfo {title} {Feynman path-integral treatment of the
  {BEC-impurity} polaron},\ }\href@noop {} {\bibfield  {journal} {\bibinfo
  {journal} {Phys.\ Rev.\ B}\ }\textbf {\bibinfo {volume} {80}},\ \bibinfo
  {pages} {184504} (\bibinfo {year} {2009})}\BibitemShut {NoStop}%
\bibitem [{\citenamefont {Casteels}\ \emph {et~al.}(2012)\citenamefont
  {Casteels}, \citenamefont {Tempere},\ and\ \citenamefont
  {Devreese}}]{casteels2012polaronic}%
  \BibitemOpen
  \bibfield  {author} {\bibinfo {author} {\bibfnamefont {W.}~\bibnamefont
  {Casteels}}, \bibinfo {author} {\bibfnamefont {J.}~\bibnamefont {Tempere}},\
  and\ \bibinfo {author} {\bibfnamefont {J.~T.}\ \bibnamefont {Devreese}},\
  }\bibfield  {title} {\bibinfo {title} {Polaronic properties of an impurity in
  a bose-einstein condensate in reduced dimensions},\ }\href@noop {} {\bibfield
   {journal} {\bibinfo  {journal} {Physical Review A}\ }\textbf {\bibinfo
  {volume} {86}},\ \bibinfo {pages} {043614} (\bibinfo {year}
  {2012})}\BibitemShut {NoStop}%
\bibitem [{\citenamefont {Grusdt}\ \emph {et~al.}(2015)\citenamefont {Grusdt},
  \citenamefont {Shchadilova}, \citenamefont {Rubtsov},\ and\ \citenamefont
  {Demler}}]{grusdt2015renormalization}%
  \BibitemOpen
  \bibfield  {author} {\bibinfo {author} {\bibfnamefont {F.}~\bibnamefont
  {Grusdt}}, \bibinfo {author} {\bibfnamefont {Y.~E.}\ \bibnamefont
  {Shchadilova}}, \bibinfo {author} {\bibfnamefont {A.~N.}\ \bibnamefont
  {Rubtsov}},\ and\ \bibinfo {author} {\bibfnamefont {E.}~\bibnamefont
  {Demler}},\ }\bibfield  {title} {\bibinfo {title} {Renormalization group
  approach to the fr{\"o}hlich polaron model: application to {impurity-BEC}
  problem},\ }\href@noop {} {\bibfield  {journal} {\bibinfo  {journal}
  {Scientific reports}\ }\textbf {\bibinfo {volume} {5}},\ \bibinfo {pages}
  {12124} (\bibinfo {year} {2015})}\BibitemShut {NoStop}%
\bibitem [{\citenamefont {Shchadilova}\ \emph {et~al.}(2016)\citenamefont
  {Shchadilova}, \citenamefont {Schmidt}, \citenamefont {Grusdt},\ and\
  \citenamefont {Demler}}]{shchadilova2016quantum}%
  \BibitemOpen
  \bibfield  {author} {\bibinfo {author} {\bibfnamefont {Y.~E.}\ \bibnamefont
  {Shchadilova}}, \bibinfo {author} {\bibfnamefont {R.}~\bibnamefont
  {Schmidt}}, \bibinfo {author} {\bibfnamefont {F.}~\bibnamefont {Grusdt}},\
  and\ \bibinfo {author} {\bibfnamefont {E.}~\bibnamefont {Demler}},\
  }\bibfield  {title} {\bibinfo {title} {Quantum dynamics of ultracold {Bose}
  polarons},\ }\href@noop {} {\bibfield  {journal} {\bibinfo  {journal} {Phys.\
  Rev.\ Lett.}\ }\textbf {\bibinfo {volume} {117}},\ \bibinfo {pages} {113002}
  (\bibinfo {year} {2016})}\BibitemShut {NoStop}%
\bibitem [{\citenamefont {Volosniev}\ and\ \citenamefont
  {Hammer}(2017)}]{volosniev2017analytical}%
  \BibitemOpen
  \bibfield  {author} {\bibinfo {author} {\bibfnamefont {A.~G.}\ \bibnamefont
  {Volosniev}}\ and\ \bibinfo {author} {\bibfnamefont {H.-W.}\ \bibnamefont
  {Hammer}},\ }\bibfield  {title} {\bibinfo {title} {Analytical approach to the
  bose-polaron problem in one dimension},\ }\href@noop {} {\bibfield  {journal}
  {\bibinfo  {journal} {Phys. Rev. A}\ }\textbf {\bibinfo {volume} {96}},\
  \bibinfo {pages} {031601} (\bibinfo {year} {2017})}\BibitemShut {NoStop}%
\bibitem [{\citenamefont {Grusdt}\ \emph {et~al.}(2017)\citenamefont {Grusdt},
  \citenamefont {Astrakharchik},\ and\ \citenamefont
  {Demler}}]{grusdt2017bose}%
  \BibitemOpen
  \bibfield  {author} {\bibinfo {author} {\bibfnamefont {F.}~\bibnamefont
  {Grusdt}}, \bibinfo {author} {\bibfnamefont {G.~E.}\ \bibnamefont
  {Astrakharchik}},\ and\ \bibinfo {author} {\bibfnamefont {E.}~\bibnamefont
  {Demler}},\ }\bibfield  {title} {\bibinfo {title} {Bose polarons in ultracold
  atoms in one dimension: beyond the fr{\"o}hlich paradigm},\ }\href@noop {}
  {\bibfield  {journal} {\bibinfo  {journal} {New Journal of Physics}\ }\textbf
  {\bibinfo {volume} {19}},\ \bibinfo {pages} {103035} (\bibinfo {year}
  {2017})}\BibitemShut {NoStop}%
\bibitem [{\citenamefont {Mistakidis}\ \emph
  {et~al.}(2019{\natexlab{a}})\citenamefont {Mistakidis}, \citenamefont
  {Volosniev}, \citenamefont {Zinner},\ and\ \citenamefont
  {Schmelcher}}]{mistakidis2019effective}%
  \BibitemOpen
  \bibfield  {author} {\bibinfo {author} {\bibfnamefont {S.~I.}\ \bibnamefont
  {Mistakidis}}, \bibinfo {author} {\bibfnamefont {A.~G.}\ \bibnamefont
  {Volosniev}}, \bibinfo {author} {\bibfnamefont {N.~T.}\ \bibnamefont
  {Zinner}},\ and\ \bibinfo {author} {\bibfnamefont {P.}~\bibnamefont
  {Schmelcher}},\ }\bibfield  {title} {\bibinfo {title} {Effective approach to
  impurity dynamics in one-dimensional trapped bose gases},\ }\href@noop {}
  {\bibfield  {journal} {\bibinfo  {journal} {Physical Review A}\ }\textbf
  {\bibinfo {volume} {100}},\ \bibinfo {pages} {013619} (\bibinfo {year}
  {2019}{\natexlab{a}})}\BibitemShut {NoStop}%
\bibitem [{\citenamefont {J{\o}rgensen}\ \emph {et~al.}(2016)\citenamefont
  {J{\o}rgensen}, \citenamefont {Wacker}, \citenamefont {Skalmstang},
  \citenamefont {Parish}, \citenamefont {Levinsen}, \citenamefont
  {Christensen}, \citenamefont {Bruun},\ and\ \citenamefont
  {Arlt}}]{jorgensen2016}%
  \BibitemOpen
  \bibfield  {author} {\bibinfo {author} {\bibfnamefont {N.~B.}\ \bibnamefont
  {J{\o}rgensen}}, \bibinfo {author} {\bibfnamefont {L.}~\bibnamefont
  {Wacker}}, \bibinfo {author} {\bibfnamefont {K.~T.}\ \bibnamefont
  {Skalmstang}}, \bibinfo {author} {\bibfnamefont {M.~M.}\ \bibnamefont
  {Parish}}, \bibinfo {author} {\bibfnamefont {J.}~\bibnamefont {Levinsen}},
  \bibinfo {author} {\bibfnamefont {R.~S.}\ \bibnamefont {Christensen}},
  \bibinfo {author} {\bibfnamefont {G.~M.}\ \bibnamefont {Bruun}},\ and\
  \bibinfo {author} {\bibfnamefont {J.~J.}\ \bibnamefont {Arlt}},\ }\bibfield
  {title} {\bibinfo {title} {Observation of attractive and repulsive polarons
  in a bose-einstein condensate},\ }\href@noop {} {\bibfield  {journal}
  {\bibinfo  {journal} {Phys.\ Rev.\ Lett.}\ }\textbf {\bibinfo {volume}
  {117}},\ \bibinfo {pages} {055302} (\bibinfo {year} {2016})}\BibitemShut
  {NoStop}%
\bibitem [{\citenamefont {Hu}\ \emph {et~al.}(2016)\citenamefont {Hu},
  \citenamefont {Van~de Graaff}, \citenamefont {Kedar}, \citenamefont {Corson},
  \citenamefont {Cornell},\ and\ \citenamefont {Jin}}]{hu2016bose}%
  \BibitemOpen
  \bibfield  {author} {\bibinfo {author} {\bibfnamefont {M.-G.}\ \bibnamefont
  {Hu}}, \bibinfo {author} {\bibfnamefont {M.~J.}\ \bibnamefont {Van~de
  Graaff}}, \bibinfo {author} {\bibfnamefont {D.}~\bibnamefont {Kedar}},
  \bibinfo {author} {\bibfnamefont {J.~P.}\ \bibnamefont {Corson}}, \bibinfo
  {author} {\bibfnamefont {E.~A.}\ \bibnamefont {Cornell}},\ and\ \bibinfo
  {author} {\bibfnamefont {D.~S.}\ \bibnamefont {Jin}},\ }\bibfield  {title}
  {\bibinfo {title} {Bose polarons in the strongly interacting regime},\
  }\href@noop {} {\bibfield  {journal} {\bibinfo  {journal} {Phys.\ Rev.\
  Lett.}\ }\textbf {\bibinfo {volume} {117}},\ \bibinfo {pages} {055301}
  (\bibinfo {year} {2016})}\BibitemShut {NoStop}%
\bibitem [{\citenamefont {Ardila}\ and\ \citenamefont
  {Giorgini}(2016)}]{ardila2016bose}%
  \BibitemOpen
  \bibfield  {author} {\bibinfo {author} {\bibfnamefont {L.~A.~P.}\
  \bibnamefont {Ardila}}\ and\ \bibinfo {author} {\bibfnamefont
  {S.}~\bibnamefont {Giorgini}},\ }\bibfield  {title} {\bibinfo {title} {Bose
  polaron problem: Effect of mass imbalance on binding energy},\ }\href@noop {}
  {\bibfield  {journal} {\bibinfo  {journal} {Physical Review A}\ }\textbf
  {\bibinfo {volume} {94}},\ \bibinfo {pages} {063640} (\bibinfo {year}
  {2016})}\BibitemShut {NoStop}%
\bibitem [{\citenamefont {Mistakidis}\ \emph
  {et~al.}(2020{\natexlab{a}})\citenamefont {Mistakidis}, \citenamefont
  {Katsimiga}, \citenamefont {Koutentakis}, \citenamefont {Busch},\ and\
  \citenamefont {Schmelcher}}]{mistakidis2020pump}%
  \BibitemOpen
  \bibfield  {author} {\bibinfo {author} {\bibfnamefont {S.~I.}\ \bibnamefont
  {Mistakidis}}, \bibinfo {author} {\bibfnamefont {G.~C.}\ \bibnamefont
  {Katsimiga}}, \bibinfo {author} {\bibfnamefont {G.~M.}\ \bibnamefont
  {Koutentakis}}, \bibinfo {author} {\bibfnamefont {T.}~\bibnamefont {Busch}},\
  and\ \bibinfo {author} {\bibfnamefont {P.}~\bibnamefont {Schmelcher}},\
  }\bibfield  {title} {\bibinfo {title} {Pump-probe spectroscopy of bose
  polarons: Dynamical formation and coherence},\ }\href@noop {} {\bibfield
  {journal} {\bibinfo  {journal} {Physical Review Research}\ }\textbf {\bibinfo
  {volume} {2}},\ \bibinfo {pages} {033380} (\bibinfo {year}
  {2020}{\natexlab{a}})}\BibitemShut {NoStop}%
\bibitem [{\citenamefont {Bruderer}\ \emph {et~al.}(2008)\citenamefont
  {Bruderer}, \citenamefont {Klein}, \citenamefont {Clark},\ and\ \citenamefont
  {Jaksch}}]{bruderer2008transport}%
  \BibitemOpen
  \bibfield  {author} {\bibinfo {author} {\bibfnamefont {M.}~\bibnamefont
  {Bruderer}}, \bibinfo {author} {\bibfnamefont {A.}~\bibnamefont {Klein}},
  \bibinfo {author} {\bibfnamefont {S.~R.}\ \bibnamefont {Clark}},\ and\
  \bibinfo {author} {\bibfnamefont {D.}~\bibnamefont {Jaksch}},\ }\bibfield
  {title} {\bibinfo {title} {Transport of strong-coupling polarons in optical
  lattices},\ }\href@noop {} {\bibfield  {journal} {\bibinfo  {journal} {New
  Journal of Physics}\ }\textbf {\bibinfo {volume} {10}},\ \bibinfo {pages}
  {033015} (\bibinfo {year} {2008})}\BibitemShut {NoStop}%
\bibitem [{\citenamefont {Cai}\ \emph {et~al.}(2010)\citenamefont {Cai},
  \citenamefont {Wang}, \citenamefont {Xie},\ and\ \citenamefont
  {Wang}}]{cai2010interaction}%
  \BibitemOpen
  \bibfield  {author} {\bibinfo {author} {\bibfnamefont {Z.}~\bibnamefont
  {Cai}}, \bibinfo {author} {\bibfnamefont {L.}~\bibnamefont {Wang}}, \bibinfo
  {author} {\bibfnamefont {X.~C.}\ \bibnamefont {Xie}},\ and\ \bibinfo {author}
  {\bibfnamefont {Y.}~\bibnamefont {Wang}},\ }\bibfield  {title} {\bibinfo
  {title} {Interaction-induced anomalous transport behavior in one-dimensional
  optical lattices},\ }\href@noop {} {\bibfield  {journal} {\bibinfo  {journal}
  {Physical Review A}\ }\textbf {\bibinfo {volume} {81}},\ \bibinfo {pages}
  {043602} (\bibinfo {year} {2010})}\BibitemShut {NoStop}%
\bibitem [{\citenamefont {Johnson}\ \emph {et~al.}(2011)\citenamefont
  {Johnson}, \citenamefont {Clark}, \citenamefont {Bruderer},\ and\
  \citenamefont {Jaksch}}]{johnson2011impurity}%
  \BibitemOpen
  \bibfield  {author} {\bibinfo {author} {\bibfnamefont {T.~H.}\ \bibnamefont
  {Johnson}}, \bibinfo {author} {\bibfnamefont {S.~R.}\ \bibnamefont {Clark}},
  \bibinfo {author} {\bibfnamefont {M.}~\bibnamefont {Bruderer}},\ and\
  \bibinfo {author} {\bibfnamefont {D.}~\bibnamefont {Jaksch}},\ }\bibfield
  {title} {\bibinfo {title} {Impurity transport through a strongly interacting
  bosonic quantum gas},\ }\href@noop {} {\bibfield  {journal} {\bibinfo
  {journal} {Physical Review A}\ }\textbf {\bibinfo {volume} {84}},\ \bibinfo
  {pages} {023617} (\bibinfo {year} {2011})}\BibitemShut {NoStop}%
\bibitem [{\citenamefont {Theel}\ \emph {et~al.}(2020)\citenamefont {Theel},
  \citenamefont {Keiler}, \citenamefont {Mistakidis},\ and\ \citenamefont
  {Schmelcher}}]{theel2020manybodycollisional}%
  \BibitemOpen
  \bibfield  {author} {\bibinfo {author} {\bibfnamefont {F.}~\bibnamefont
  {Theel}}, \bibinfo {author} {\bibfnamefont {K.}~\bibnamefont {Keiler}},
  \bibinfo {author} {\bibfnamefont {S.~I.}\ \bibnamefont {Mistakidis}},\ and\
  \bibinfo {author} {\bibfnamefont {P.}~\bibnamefont {Schmelcher}},\ }\bibfield
   {title} {\bibinfo {title} {Many-body collisional dynamics of impurities
  injected into a double-well trapped bose-einstein condensate},\ }\href@noop
  {} {\bibfield  {journal} {\bibinfo  {journal} {arXiv preprint
  arXiv:2009.12147}\ } (\bibinfo {year} {2020})}\BibitemShut {NoStop}%
\bibitem [{\citenamefont {Palzer}\ \emph {et~al.}(2009)\citenamefont {Palzer},
  \citenamefont {Zipkes}, \citenamefont {Sias},\ and\ \citenamefont
  {K{\"o}hl}}]{palzer2009quantum}%
  \BibitemOpen
  \bibfield  {author} {\bibinfo {author} {\bibfnamefont {S.}~\bibnamefont
  {Palzer}}, \bibinfo {author} {\bibfnamefont {C.}~\bibnamefont {Zipkes}},
  \bibinfo {author} {\bibfnamefont {C.}~\bibnamefont {Sias}},\ and\ \bibinfo
  {author} {\bibfnamefont {M.}~\bibnamefont {K{\"o}hl}},\ }\bibfield  {title}
  {\bibinfo {title} {Quantum transport through a tonks-girardeau gas},\
  }\href@noop {} {\bibfield  {journal} {\bibinfo  {journal} {Physical review
  letters}\ }\textbf {\bibinfo {volume} {103}},\ \bibinfo {pages} {150601}
  (\bibinfo {year} {2009})}\BibitemShut {NoStop}%
\bibitem [{\citenamefont {Frese}\ \emph {et~al.}(2000)\citenamefont {Frese},
  \citenamefont {Ueberholz}, \citenamefont {Kuhr}, \citenamefont {Alt},
  \citenamefont {Schrader}, \citenamefont {Gomer},\ and\ \citenamefont
  {Meschede}}]{frese2000single}%
  \BibitemOpen
  \bibfield  {author} {\bibinfo {author} {\bibfnamefont {D.}~\bibnamefont
  {Frese}}, \bibinfo {author} {\bibfnamefont {B.}~\bibnamefont {Ueberholz}},
  \bibinfo {author} {\bibfnamefont {S.}~\bibnamefont {Kuhr}}, \bibinfo {author}
  {\bibfnamefont {W.}~\bibnamefont {Alt}}, \bibinfo {author} {\bibfnamefont
  {D.}~\bibnamefont {Schrader}}, \bibinfo {author} {\bibfnamefont
  {V.}~\bibnamefont {Gomer}},\ and\ \bibinfo {author} {\bibfnamefont
  {D.}~\bibnamefont {Meschede}},\ }\bibfield  {title} {\bibinfo {title} {Single
  atoms in an optical dipole trap: Towards a deterministic source of cold
  atoms},\ }\href@noop {} {\bibfield  {journal} {\bibinfo  {journal} {Physical
  review letters}\ }\textbf {\bibinfo {volume} {85}},\ \bibinfo {pages} {3777}
  (\bibinfo {year} {2000})}\BibitemShut {NoStop}%
\bibitem [{\citenamefont {Bernien}\ \emph {et~al.}(2017)\citenamefont
  {Bernien}, \citenamefont {Schwartz}, \citenamefont {Keesling}, \citenamefont
  {Levine}, \citenamefont {Omran}, \citenamefont {Pichler}, \citenamefont
  {Choi}, \citenamefont {Zibrov}, \citenamefont {Endres}, \citenamefont
  {Greiner} \emph {et~al.}}]{bernien2017probing}%
  \BibitemOpen
  \bibfield  {author} {\bibinfo {author} {\bibfnamefont {H.}~\bibnamefont
  {Bernien}}, \bibinfo {author} {\bibfnamefont {S.}~\bibnamefont {Schwartz}},
  \bibinfo {author} {\bibfnamefont {A.}~\bibnamefont {Keesling}}, \bibinfo
  {author} {\bibfnamefont {H.}~\bibnamefont {Levine}}, \bibinfo {author}
  {\bibfnamefont {A.}~\bibnamefont {Omran}}, \bibinfo {author} {\bibfnamefont
  {H.}~\bibnamefont {Pichler}}, \bibinfo {author} {\bibfnamefont
  {S.}~\bibnamefont {Choi}}, \bibinfo {author} {\bibfnamefont {A.~S.}\
  \bibnamefont {Zibrov}}, \bibinfo {author} {\bibfnamefont {M.}~\bibnamefont
  {Endres}}, \bibinfo {author} {\bibfnamefont {M.}~\bibnamefont {Greiner}},
  \emph {et~al.},\ }\bibfield  {title} {\bibinfo {title} {Probing many-body
  dynamics on a 51-atom quantum simulator},\ }\href@noop {} {\bibfield
  {journal} {\bibinfo  {journal} {Nature}\ }\textbf {\bibinfo {volume} {551}},\
  \bibinfo {pages} {579} (\bibinfo {year} {2017})}\BibitemShut {NoStop}%
\bibitem [{\citenamefont {Stauber}\ \emph {et~al.}(2000)\citenamefont
  {Stauber}, \citenamefont {Zimmermann},\ and\ \citenamefont
  {Castella}}]{stauber2000electron}%
  \BibitemOpen
  \bibfield  {author} {\bibinfo {author} {\bibfnamefont {T.}~\bibnamefont
  {Stauber}}, \bibinfo {author} {\bibfnamefont {R.}~\bibnamefont
  {Zimmermann}},\ and\ \bibinfo {author} {\bibfnamefont {H.}~\bibnamefont
  {Castella}},\ }\bibfield  {title} {\bibinfo {title} {Electron-phonon
  interaction in quantum dots: A solvable model},\ }\href@noop {} {\bibfield
  {journal} {\bibinfo  {journal} {Physical Review B}\ }\textbf {\bibinfo
  {volume} {62}},\ \bibinfo {pages} {7336} (\bibinfo {year}
  {2000})}\BibitemShut {NoStop}%
\bibitem [{\citenamefont {Loss}\ and\ \citenamefont
  {DiVincenzo}(1998)}]{loss1998quantum}%
  \BibitemOpen
  \bibfield  {author} {\bibinfo {author} {\bibfnamefont {D.}~\bibnamefont
  {Loss}}\ and\ \bibinfo {author} {\bibfnamefont {D.~P.}\ \bibnamefont
  {DiVincenzo}},\ }\bibfield  {title} {\bibinfo {title} {Quantum computation
  with quantum dots},\ }\href@noop {} {\bibfield  {journal} {\bibinfo
  {journal} {Physical Review A}\ ,\ \bibinfo {pages} {120}} (\bibinfo {year}
  {1998})}\BibitemShut {NoStop}%
\bibitem [{\citenamefont {Weiss}\ \emph {et~al.}(1987)\citenamefont {Weiss},
  \citenamefont {Grabert}, \citenamefont {H{\"a}nggi},\ and\ \citenamefont
  {Riseborough}}]{weiss1987incoherent}%
  \BibitemOpen
  \bibfield  {author} {\bibinfo {author} {\bibfnamefont {U.}~\bibnamefont
  {Weiss}}, \bibinfo {author} {\bibfnamefont {H.}~\bibnamefont {Grabert}},
  \bibinfo {author} {\bibfnamefont {P.}~\bibnamefont {H{\"a}nggi}},\ and\
  \bibinfo {author} {\bibfnamefont {P.}~\bibnamefont {Riseborough}},\
  }\bibfield  {title} {\bibinfo {title} {Incoherent tunneling in a double
  well},\ }\href@noop {} {\bibfield  {journal} {\bibinfo  {journal} {Physical
  Review B}\ }\textbf {\bibinfo {volume} {35}},\ \bibinfo {pages} {9535}
  (\bibinfo {year} {1987})}\BibitemShut {NoStop}%
\bibitem [{\citenamefont {Isakov}\ \emph {et~al.}(2016)\citenamefont {Isakov},
  \citenamefont {Mazzola}, \citenamefont {Smelyanskiy}, \citenamefont {Jiang},
  \citenamefont {Boixo}, \citenamefont {Neven},\ and\ \citenamefont
  {Troyer}}]{isakov2016understanding}%
  \BibitemOpen
  \bibfield  {author} {\bibinfo {author} {\bibfnamefont {S.~V.}\ \bibnamefont
  {Isakov}}, \bibinfo {author} {\bibfnamefont {G.}~\bibnamefont {Mazzola}},
  \bibinfo {author} {\bibfnamefont {V.~N.}\ \bibnamefont {Smelyanskiy}},
  \bibinfo {author} {\bibfnamefont {Z.}~\bibnamefont {Jiang}}, \bibinfo
  {author} {\bibfnamefont {S.}~\bibnamefont {Boixo}}, \bibinfo {author}
  {\bibfnamefont {H.}~\bibnamefont {Neven}},\ and\ \bibinfo {author}
  {\bibfnamefont {M.}~\bibnamefont {Troyer}},\ }\bibfield  {title} {\bibinfo
  {title} {Understanding quantum tunneling through quantum monte carlo
  simulations},\ }\href@noop {} {\bibfield  {journal} {\bibinfo  {journal}
  {Physical review letters}\ }\textbf {\bibinfo {volume} {117}},\ \bibinfo
  {pages} {180402} (\bibinfo {year} {2016})}\BibitemShut {NoStop}%
\bibitem [{\citenamefont {Lingua}\ \emph {et~al.}(2018)\citenamefont {Lingua},
  \citenamefont {Capogrosso-Sansone}, \citenamefont {Safavi-Naini},
  \citenamefont {Jahangiri},\ and\ \citenamefont
  {Penna}}]{lingua2018multiworm}%
  \BibitemOpen
  \bibfield  {author} {\bibinfo {author} {\bibfnamefont {F.}~\bibnamefont
  {Lingua}}, \bibinfo {author} {\bibfnamefont {B.}~\bibnamefont
  {Capogrosso-Sansone}}, \bibinfo {author} {\bibfnamefont {A.}~\bibnamefont
  {Safavi-Naini}}, \bibinfo {author} {\bibfnamefont {A.~J.}\ \bibnamefont
  {Jahangiri}},\ and\ \bibinfo {author} {\bibfnamefont {V.}~\bibnamefont
  {Penna}},\ }\bibfield  {title} {\bibinfo {title} {Multiworm algorithm quantum
  monte carlo},\ }\href@noop {} {\bibfield  {journal} {\bibinfo  {journal}
  {Physica Scripta}\ }\textbf {\bibinfo {volume} {93}},\ \bibinfo {pages}
  {105402} (\bibinfo {year} {2018})}\BibitemShut {NoStop}%
\bibitem [{\citenamefont {Ardila}\ and\ \citenamefont
  {Giorgini}(2015)}]{ardila2015impurity}%
  \BibitemOpen
  \bibfield  {author} {\bibinfo {author} {\bibfnamefont {L.~A.~P.}\
  \bibnamefont {Ardila}}\ and\ \bibinfo {author} {\bibfnamefont
  {S.}~\bibnamefont {Giorgini}},\ }\bibfield  {title} {\bibinfo {title}
  {Impurity in a bose-einstein condensate: Study of the attractive and
  repulsive branch using quantum monte carlo methods},\ }\href@noop {}
  {\bibfield  {journal} {\bibinfo  {journal} {Physical Review A}\ }\textbf
  {\bibinfo {volume} {92}},\ \bibinfo {pages} {033612} (\bibinfo {year}
  {2015})}\BibitemShut {NoStop}%
\bibitem [{\citenamefont {Inack}\ and\ \citenamefont
  {Pilati}(2015)}]{inack2015simulated}%
  \BibitemOpen
  \bibfield  {author} {\bibinfo {author} {\bibfnamefont {E.~M.}\ \bibnamefont
  {Inack}}\ and\ \bibinfo {author} {\bibfnamefont {S.}~\bibnamefont {Pilati}},\
  }\bibfield  {title} {\bibinfo {title} {Simulated quantum annealing of
  double-well and multiwell potentials},\ }\href@noop {} {\bibfield  {journal}
  {\bibinfo  {journal} {Physical Review E}\ }\textbf {\bibinfo {volume} {92}},\
  \bibinfo {pages} {053304} (\bibinfo {year} {2015})}\BibitemShut {NoStop}%
\bibitem [{\citenamefont {Stella}\ \emph {et~al.}(2006)\citenamefont {Stella},
  \citenamefont {Santoro},\ and\ \citenamefont {Tosatti}}]{stella2006monte}%
  \BibitemOpen
  \bibfield  {author} {\bibinfo {author} {\bibfnamefont {L.}~\bibnamefont
  {Stella}}, \bibinfo {author} {\bibfnamefont {G.~E.}\ \bibnamefont
  {Santoro}},\ and\ \bibinfo {author} {\bibfnamefont {E.}~\bibnamefont
  {Tosatti}},\ }\bibfield  {title} {\bibinfo {title} {Monte carlo studies of
  quantum and classical annealing on a double well},\ }\href@noop {} {\bibfield
   {journal} {\bibinfo  {journal} {Physical Review B}\ }\textbf {\bibinfo
  {volume} {73}},\ \bibinfo {pages} {144302} (\bibinfo {year}
  {2006})}\BibitemShut {NoStop}%
\bibitem [{\citenamefont {Ardila}\ \emph {et~al.}(2019)\citenamefont {Ardila},
  \citenamefont {J{\o}rgensen}, \citenamefont {Pohl}, \citenamefont {Giorgini},
  \citenamefont {Bruun},\ and\ \citenamefont {Arlt}}]{ardila2019analyzing}%
  \BibitemOpen
  \bibfield  {author} {\bibinfo {author} {\bibfnamefont {L.~A.~P.}\
  \bibnamefont {Ardila}}, \bibinfo {author} {\bibfnamefont {N.~B.}\
  \bibnamefont {J{\o}rgensen}}, \bibinfo {author} {\bibfnamefont
  {T.}~\bibnamefont {Pohl}}, \bibinfo {author} {\bibfnamefont {S.}~\bibnamefont
  {Giorgini}}, \bibinfo {author} {\bibfnamefont {G.~M.}\ \bibnamefont
  {Bruun}},\ and\ \bibinfo {author} {\bibfnamefont {J.~J.}\ \bibnamefont
  {Arlt}},\ }\bibfield  {title} {\bibinfo {title} {Analyzing a bose polaron
  across resonant interactions},\ }\href@noop {} {\bibfield  {journal}
  {\bibinfo  {journal} {Physical Review A}\ }\textbf {\bibinfo {volume} {99}},\
  \bibinfo {pages} {063607} (\bibinfo {year} {2019})}\BibitemShut {NoStop}%
\bibitem [{\citenamefont {Gull}\ \emph {et~al.}(2011)\citenamefont {Gull},
  \citenamefont {Millis}, \citenamefont {Lichtenstein}, \citenamefont
  {Rubtsov}, \citenamefont {Troyer},\ and\ \citenamefont
  {Werner}}]{gull2011continuous}%
  \BibitemOpen
  \bibfield  {author} {\bibinfo {author} {\bibfnamefont {E.}~\bibnamefont
  {Gull}}, \bibinfo {author} {\bibfnamefont {A.~J.}\ \bibnamefont {Millis}},
  \bibinfo {author} {\bibfnamefont {A.~I.}\ \bibnamefont {Lichtenstein}},
  \bibinfo {author} {\bibfnamefont {A.~N.}\ \bibnamefont {Rubtsov}}, \bibinfo
  {author} {\bibfnamefont {M.}~\bibnamefont {Troyer}},\ and\ \bibinfo {author}
  {\bibfnamefont {P.}~\bibnamefont {Werner}},\ }\bibfield  {title} {\bibinfo
  {title} {Continuous-time monte carlo methods for quantum impurity models},\
  }\href@noop {} {\bibfield  {journal} {\bibinfo  {journal} {Reviews of Modern
  Physics}\ }\textbf {\bibinfo {volume} {83}},\ \bibinfo {pages} {349}
  (\bibinfo {year} {2011})}\BibitemShut {NoStop}%
\bibitem [{\citenamefont {Inack}\ \emph {et~al.}(2018)\citenamefont {Inack},
  \citenamefont {Giudici}, \citenamefont {Parolini}, \citenamefont {Santoro},\
  and\ \citenamefont {Pilati}}]{inack2018understanding}%
  \BibitemOpen
  \bibfield  {author} {\bibinfo {author} {\bibfnamefont {E.~M.}\ \bibnamefont
  {Inack}}, \bibinfo {author} {\bibfnamefont {G.}~\bibnamefont {Giudici}},
  \bibinfo {author} {\bibfnamefont {T.}~\bibnamefont {Parolini}}, \bibinfo
  {author} {\bibfnamefont {G.}~\bibnamefont {Santoro}},\ and\ \bibinfo {author}
  {\bibfnamefont {S.}~\bibnamefont {Pilati}},\ }\bibfield  {title} {\bibinfo
  {title} {Understanding quantum tunneling using diffusion monte carlo
  simulations},\ }\href@noop {} {\bibfield  {journal} {\bibinfo  {journal}
  {Physical Review A}\ }\textbf {\bibinfo {volume} {97}},\ \bibinfo {pages}
  {032307} (\bibinfo {year} {2018})}\BibitemShut {NoStop}%
\bibitem [{\citenamefont {Nemec}(2010)}]{nemec2010diffusion}%
  \BibitemOpen
  \bibfield  {author} {\bibinfo {author} {\bibfnamefont {N.}~\bibnamefont
  {Nemec}},\ }\bibfield  {title} {\bibinfo {title} {Diffusion monte carlo:
  Exponential scaling of computational cost for large systems},\ }\href@noop {}
  {\bibfield  {journal} {\bibinfo  {journal} {Physical Review B}\ }\textbf
  {\bibinfo {volume} {81}},\ \bibinfo {pages} {035119} (\bibinfo {year}
  {2010})}\BibitemShut {NoStop}%
\bibitem [{\citenamefont {Oberli}\ \emph {et~al.}(1990)\citenamefont {Oberli},
  \citenamefont {Shah}, \citenamefont {Damen}, \citenamefont {Kuo},
  \citenamefont {Henry}, \citenamefont {Lary},\ and\ \citenamefont
  {Goodnick}}]{oberli1990optical}%
  \BibitemOpen
  \bibfield  {author} {\bibinfo {author} {\bibfnamefont {D.~Y.}\ \bibnamefont
  {Oberli}}, \bibinfo {author} {\bibfnamefont {J.}~\bibnamefont {Shah}},
  \bibinfo {author} {\bibfnamefont {T.~C.}\ \bibnamefont {Damen}}, \bibinfo
  {author} {\bibfnamefont {J.~M.}\ \bibnamefont {Kuo}}, \bibinfo {author}
  {\bibfnamefont {J.~E.}\ \bibnamefont {Henry}}, \bibinfo {author}
  {\bibfnamefont {J.}~\bibnamefont {Lary}},\ and\ \bibinfo {author}
  {\bibfnamefont {S.~M.}\ \bibnamefont {Goodnick}},\ }\bibfield  {title}
  {\bibinfo {title} {Optical phonon-assisted tunneling in double quantum well
  structures},\ }\href@noop {} {\bibfield  {journal} {\bibinfo  {journal}
  {Applied physics letters}\ }\textbf {\bibinfo {volume} {56}},\ \bibinfo
  {pages} {1239} (\bibinfo {year} {1990})}\BibitemShut {NoStop}%
\bibitem [{\citenamefont {Vargas-Calder{\'o}n}\ and\ \citenamefont
  {Vinck-Posada}(2020)}]{vargas2020light}%
  \BibitemOpen
  \bibfield  {author} {\bibinfo {author} {\bibfnamefont {V.}~\bibnamefont
  {Vargas-Calder{\'o}n}}\ and\ \bibinfo {author} {\bibfnamefont
  {H.}~\bibnamefont {Vinck-Posada}},\ }\bibfield  {title} {\bibinfo {title}
  {Light emission properties in a double quantum dot molecule immersed in a
  cavity: phonon-assisted tunneling},\ }\href@noop {} {\bibfield  {journal}
  {\bibinfo  {journal} {Physics Letters A}\ }\textbf {\bibinfo {volume}
  {384}},\ \bibinfo {pages} {126076} (\bibinfo {year} {2020})}\BibitemShut
  {NoStop}%
\bibitem [{\citenamefont {Myasnikova}(1995)}]{myasnikova1995band}%
  \BibitemOpen
  \bibfield  {author} {\bibinfo {author} {\bibfnamefont {A.~E.}\ \bibnamefont
  {Myasnikova}},\ }\bibfield  {title} {\bibinfo {title} {Band structure in
  autolocalization and bipolaron models of high-temperature
  superconductivity},\ }\href@noop {} {\bibfield  {journal} {\bibinfo
  {journal} {Physical Review B}\ }\textbf {\bibinfo {volume} {52}},\ \bibinfo
  {pages} {10457} (\bibinfo {year} {1995})}\BibitemShut {NoStop}%
\bibitem [{\citenamefont {Westbroek}\ \emph {et~al.}(2018)\citenamefont
  {Westbroek}, \citenamefont {King}, \citenamefont {Vvedensky},\ and\
  \citenamefont {D{\"u}rr}}]{westbroek2018user}%
  \BibitemOpen
  \bibfield  {author} {\bibinfo {author} {\bibfnamefont {M.~J.~E.}\
  \bibnamefont {Westbroek}}, \bibinfo {author} {\bibfnamefont {P.~R.}\
  \bibnamefont {King}}, \bibinfo {author} {\bibfnamefont {D.~D.}\ \bibnamefont
  {Vvedensky}},\ and\ \bibinfo {author} {\bibfnamefont {S.}~\bibnamefont
  {D{\"u}rr}},\ }\bibfield  {title} {\bibinfo {title} {User's guide to monte
  carlo methods for evaluating path integrals},\ }\href@noop {} {\bibfield
  {journal} {\bibinfo  {journal} {American Journal of Physics}\ }\textbf
  {\bibinfo {volume} {86}},\ \bibinfo {pages} {293} (\bibinfo {year}
  {2018})}\BibitemShut {NoStop}%
\bibitem [{\citenamefont {Metropolis}\ and\ \citenamefont
  {Ulam}(1949)}]{metropolis1949monte}%
  \BibitemOpen
  \bibfield  {author} {\bibinfo {author} {\bibfnamefont {N.}~\bibnamefont
  {Metropolis}}\ and\ \bibinfo {author} {\bibfnamefont {S.}~\bibnamefont
  {Ulam}},\ }\bibfield  {title} {\bibinfo {title} {The monte carlo method},\
  }\href@noop {} {\bibfield  {journal} {\bibinfo  {journal} {Journal of the
  American statistical association}\ }\textbf {\bibinfo {volume} {44}},\
  \bibinfo {pages} {335} (\bibinfo {year} {1949})}\BibitemShut {NoStop}%
\bibitem [{\citenamefont {Parolini}\ \emph {et~al.}(2019)\citenamefont
  {Parolini}, \citenamefont {Inack}, \citenamefont {Giudici},\ and\
  \citenamefont {Pilati}}]{parolini2019tunneling}%
  \BibitemOpen
  \bibfield  {author} {\bibinfo {author} {\bibfnamefont {T.}~\bibnamefont
  {Parolini}}, \bibinfo {author} {\bibfnamefont {E.~M.}\ \bibnamefont {Inack}},
  \bibinfo {author} {\bibfnamefont {G.}~\bibnamefont {Giudici}},\ and\ \bibinfo
  {author} {\bibfnamefont {S.}~\bibnamefont {Pilati}},\ }\bibfield  {title}
  {\bibinfo {title} {Tunneling in projective quantum monte carlo simulations
  with guiding wave functions},\ }\href@noop {} {\bibfield  {journal} {\bibinfo
   {journal} {Physical Review B}\ }\textbf {\bibinfo {volume} {100}},\ \bibinfo
  {pages} {214303} (\bibinfo {year} {2019})}\BibitemShut {NoStop}%
\bibitem [{\citenamefont {Feynman}\ \emph {et~al.}(2010)\citenamefont
  {Feynman}, \citenamefont {Hibbs},\ and\ \citenamefont
  {Styer}}]{feynman2010quantum}%
  \BibitemOpen
  \bibfield  {author} {\bibinfo {author} {\bibfnamefont {R.~P.}\ \bibnamefont
  {Feynman}}, \bibinfo {author} {\bibfnamefont {A.~R.}\ \bibnamefont {Hibbs}},\
  and\ \bibinfo {author} {\bibfnamefont {D.~F.}\ \bibnamefont {Styer}},\
  }\href@noop {} {\emph {\bibinfo {title} {Quantum mechanics and path
  integrals}}}\ (\bibinfo  {publisher} {Courier Corporation},\ \bibinfo {year}
  {2010})\BibitemShut {NoStop}%
\bibitem [{\citenamefont {Levy}\ \emph {et~al.}(2017)\citenamefont {Levy},
  \citenamefont {LeBlanc},\ and\ \citenamefont
  {Gull}}]{levy2017implementation}%
  \BibitemOpen
  \bibfield  {author} {\bibinfo {author} {\bibfnamefont {R.}~\bibnamefont
  {Levy}}, \bibinfo {author} {\bibfnamefont {J.~P.~F.}\ \bibnamefont
  {LeBlanc}},\ and\ \bibinfo {author} {\bibfnamefont {E.}~\bibnamefont
  {Gull}},\ }\bibfield  {title} {\bibinfo {title} {Implementation of the
  maximum entropy method for analytic continuation},\ }\href@noop {} {\bibfield
   {journal} {\bibinfo  {journal} {Computer Physics Communications}\ }\textbf
  {\bibinfo {volume} {215}},\ \bibinfo {pages} {149} (\bibinfo {year}
  {2017})}\BibitemShut {NoStop}%
\bibitem [{\citenamefont {Ghanem}\ and\ \citenamefont
  {Koch}(2016)}]{ghanem2016analytic}%
  \BibitemOpen
  \bibfield  {author} {\bibinfo {author} {\bibfnamefont {K.}~\bibnamefont
  {Ghanem}}\ and\ \bibinfo {author} {\bibfnamefont {E.}~\bibnamefont {Koch}},\
  }\bibfield  {title} {\bibinfo {title} {Analytic continuation of quantum monte
  carlo data. stochastic sampling method},\ }\href@noop {} {\bibfield
  {journal} {\bibinfo  {journal} {Verhandlungen der Deutschen Physikalischen
  Gesellschaft}\ } (\bibinfo {year} {2016})}\BibitemShut {NoStop}%
\bibitem [{\citenamefont {Pilati}\ \emph {et~al.}(2006)\citenamefont {Pilati},
  \citenamefont {Sakkos}, \citenamefont {Boronat}, \citenamefont {Casulleras},\
  and\ \citenamefont {Giorgini}}]{pilati2006equation}%
  \BibitemOpen
  \bibfield  {author} {\bibinfo {author} {\bibfnamefont {S.}~\bibnamefont
  {Pilati}}, \bibinfo {author} {\bibfnamefont {K.}~\bibnamefont {Sakkos}},
  \bibinfo {author} {\bibfnamefont {J.}~\bibnamefont {Boronat}}, \bibinfo
  {author} {\bibfnamefont {J.}~\bibnamefont {Casulleras}},\ and\ \bibinfo
  {author} {\bibfnamefont {S.}~\bibnamefont {Giorgini}},\ }\bibfield  {title}
  {\bibinfo {title} {Equation of state of an interacting bose gas at finite
  temperature: A path-integral monte carlo study},\ }\href@noop {} {\bibfield
  {journal} {\bibinfo  {journal} {Physical Review A}\ }\textbf {\bibinfo
  {volume} {74}},\ \bibinfo {pages} {043621} (\bibinfo {year}
  {2006})}\BibitemShut {NoStop}%
\bibitem [{\citenamefont {Ceperley}(1995)}]{ceperley1995path}%
  \BibitemOpen
  \bibfield  {author} {\bibinfo {author} {\bibfnamefont {D.~M.}\ \bibnamefont
  {Ceperley}},\ }\bibfield  {title} {\bibinfo {title} {Path integrals in the
  theory of condensed helium},\ }\href@noop {} {\bibfield  {journal} {\bibinfo
  {journal} {Reviews of Modern Physics}\ }\textbf {\bibinfo {volume} {67}},\
  \bibinfo {pages} {279} (\bibinfo {year} {1995})}\BibitemShut {NoStop}%
\bibitem [{\citenamefont {Krauth}(1996)}]{krauth1996quantum}%
  \BibitemOpen
  \bibfield  {author} {\bibinfo {author} {\bibfnamefont {W.}~\bibnamefont
  {Krauth}},\ }\bibfield  {title} {\bibinfo {title} {Quantum monte carlo
  calculations for a large number of bosons in a harmonic trap},\ }\href@noop
  {} {\bibfield  {journal} {\bibinfo  {journal} {Physical review letters}\
  }\textbf {\bibinfo {volume} {77}},\ \bibinfo {pages} {3695} (\bibinfo {year}
  {1996})}\BibitemShut {NoStop}%
\bibitem [{\citenamefont {Yoon}\ \emph {et~al.}(2018)\citenamefont {Yoon},
  \citenamefont {Sim},\ and\ \citenamefont {Han}}]{yoon2018analytic}%
  \BibitemOpen
  \bibfield  {author} {\bibinfo {author} {\bibfnamefont {H.}~\bibnamefont
  {Yoon}}, \bibinfo {author} {\bibfnamefont {J.-H.}\ \bibnamefont {Sim}},\ and\
  \bibinfo {author} {\bibfnamefont {M.~J.}\ \bibnamefont {Han}},\ }\bibfield
  {title} {\bibinfo {title} {Analytic continuation via domain knowledge free
  machine learning},\ }\href@noop {} {\bibfield  {journal} {\bibinfo  {journal}
  {Physical Review B}\ }\textbf {\bibinfo {volume} {98}},\ \bibinfo {pages}
  {245101} (\bibinfo {year} {2018})}\BibitemShut {NoStop}%
\bibitem [{\citenamefont {Mead}\ and\ \citenamefont
  {Papanicolaou}(1984)}]{mead1984maximum}%
  \BibitemOpen
  \bibfield  {author} {\bibinfo {author} {\bibfnamefont {L.~R.}\ \bibnamefont
  {Mead}}\ and\ \bibinfo {author} {\bibfnamefont {N.}~\bibnamefont
  {Papanicolaou}},\ }\bibfield  {title} {\bibinfo {title} {Maximum entropy in
  the problem of moments},\ }\href@noop {} {\bibfield  {journal} {\bibinfo
  {journal} {Journal of Mathematical Physics}\ }\textbf {\bibinfo {volume}
  {25}},\ \bibinfo {pages} {2404} (\bibinfo {year} {1984})}\BibitemShut
  {NoStop}%
\bibitem [{\citenamefont {Spethmann}\ \emph {et~al.}(2012)\citenamefont
  {Spethmann}, \citenamefont {Kindermann}, \citenamefont {John}, \citenamefont
  {Weber}, \citenamefont {Meschede},\ and\ \citenamefont
  {Widera}}]{spethmann2012dynamics}%
  \BibitemOpen
  \bibfield  {author} {\bibinfo {author} {\bibfnamefont {N.}~\bibnamefont
  {Spethmann}}, \bibinfo {author} {\bibfnamefont {F.}~\bibnamefont
  {Kindermann}}, \bibinfo {author} {\bibfnamefont {S.}~\bibnamefont {John}},
  \bibinfo {author} {\bibfnamefont {C.}~\bibnamefont {Weber}}, \bibinfo
  {author} {\bibfnamefont {D.}~\bibnamefont {Meschede}},\ and\ \bibinfo
  {author} {\bibfnamefont {A.}~\bibnamefont {Widera}},\ }\bibfield  {title}
  {\bibinfo {title} {Dynamics of single neutral impurity atoms immersed in an
  ultracold gas},\ }\href@noop {} {\bibfield  {journal} {\bibinfo  {journal}
  {Phys.\ Rev.\ Lett.}\ }\textbf {\bibinfo {volume} {109}},\ \bibinfo {pages}
  {235301} (\bibinfo {year} {2012})}\BibitemShut {NoStop}%
\bibitem [{\citenamefont {Catani}\ \emph {et~al.}(2012)\citenamefont {Catani},
  \citenamefont {Lamporesi}, \citenamefont {Naik}, \citenamefont {Gring},
  \citenamefont {Inguscio}, \citenamefont {Minardi}, \citenamefont {Kantian},\
  and\ \citenamefont {Giamarchi}}]{catani2012quantum}%
  \BibitemOpen
  \bibfield  {author} {\bibinfo {author} {\bibfnamefont {J.}~\bibnamefont
  {Catani}}, \bibinfo {author} {\bibfnamefont {G.}~\bibnamefont {Lamporesi}},
  \bibinfo {author} {\bibfnamefont {D.}~\bibnamefont {Naik}}, \bibinfo {author}
  {\bibfnamefont {M.}~\bibnamefont {Gring}}, \bibinfo {author} {\bibfnamefont
  {M.}~\bibnamefont {Inguscio}}, \bibinfo {author} {\bibfnamefont
  {F.}~\bibnamefont {Minardi}}, \bibinfo {author} {\bibfnamefont
  {A.}~\bibnamefont {Kantian}},\ and\ \bibinfo {author} {\bibfnamefont
  {T.}~\bibnamefont {Giamarchi}},\ }\bibfield  {title} {\bibinfo {title}
  {Quantum dynamics of impurities in a one-dimensional bose gas},\ }\href@noop
  {} {\bibfield  {journal} {\bibinfo  {journal} {Physical Review A}\ }\textbf
  {\bibinfo {volume} {85}},\ \bibinfo {pages} {023623} (\bibinfo {year}
  {2012})}\BibitemShut {NoStop}%
\bibitem [{\citenamefont {Mistakidis}\ \emph
  {et~al.}(2020{\natexlab{b}})\citenamefont {Mistakidis}, \citenamefont
  {Koutentakis}, \citenamefont {Katsimiga}, \citenamefont {Busch},\ and\
  \citenamefont {Schmelcher}}]{mistakidis2020many}%
  \BibitemOpen
  \bibfield  {author} {\bibinfo {author} {\bibfnamefont {S.~I.}\ \bibnamefont
  {Mistakidis}}, \bibinfo {author} {\bibfnamefont {G.~M.}\ \bibnamefont
  {Koutentakis}}, \bibinfo {author} {\bibfnamefont {G.~C.}\ \bibnamefont
  {Katsimiga}}, \bibinfo {author} {\bibfnamefont {T.}~\bibnamefont {Busch}},\
  and\ \bibinfo {author} {\bibfnamefont {P.}~\bibnamefont {Schmelcher}},\
  }\bibfield  {title} {\bibinfo {title} {Many-body quantum dynamics and induced
  correlations of bose polarons},\ }\href@noop {} {\bibfield  {journal}
  {\bibinfo  {journal} {New Journal of Physics}\ }\textbf {\bibinfo {volume}
  {22}},\ \bibinfo {pages} {043007} (\bibinfo {year}
  {2020}{\natexlab{b}})}\BibitemShut {NoStop}%
\bibitem [{\citenamefont {Mistakidis}\ \emph
  {et~al.}(2019{\natexlab{b}})\citenamefont {Mistakidis}, \citenamefont
  {Katsimiga}, \citenamefont {Koutentakis}, \citenamefont {Busch},\ and\
  \citenamefont {Schmelcher}}]{mistakidis2019quench}%
  \BibitemOpen
  \bibfield  {author} {\bibinfo {author} {\bibfnamefont {S.~I.}\ \bibnamefont
  {Mistakidis}}, \bibinfo {author} {\bibfnamefont {G.~C.}\ \bibnamefont
  {Katsimiga}}, \bibinfo {author} {\bibfnamefont {G.~M.}\ \bibnamefont
  {Koutentakis}}, \bibinfo {author} {\bibfnamefont {T.}~\bibnamefont {Busch}},\
  and\ \bibinfo {author} {\bibfnamefont {P.}~\bibnamefont {Schmelcher}},\
  }\bibfield  {title} {\bibinfo {title} {Quench dynamics and orthogonality
  catastrophe of bose polarons},\ }\href@noop {} {\bibfield  {journal}
  {\bibinfo  {journal} {Physical review letters}\ }\textbf {\bibinfo {volume}
  {122}},\ \bibinfo {pages} {183001} (\bibinfo {year}
  {2019}{\natexlab{b}})}\BibitemShut {NoStop}%
\end{thebibliography}%

\end{document}